\documentclass[10pt,letterpaper]{article}
\usepackage{ccn}
\usepackage{pslatex}
\usepackage{apacite}
\usepackage{amsmath}
\usepackage{amsfonts}
\usepackage{bm}
\usepackage{graphicx}
\usepackage{booktabs}

\usepackage[ruled,vlined]{algorithm2e}

\usepackage{amssymb}
\usepackage{color}

\title{Probabilistic Successor Representations with Kalman Temporal Differences}

\author{{\large \bf Jesse P. Geerts\textsuperscript{1, 2}, Kimberly L. Stachenfeld\textsuperscript{3}, Neil Burgess\textsuperscript{2} } \\
  {\bf jesse.geerts.14@ucl.ac.uk, stachenfeld@google.com, n.burgess@ucl.ac.uk} \\
  \textsuperscript{1}Sainsbury Wellcome Centre, \textsuperscript{2} Institute of Cognitive Neuroscience, University College London; \textsuperscript{3}DeepMind \\ London, UK
}

\DeclareMathOperator{\E}{\mathbb{E}}
\DeclareMathOperator{\I}{\mathbb{I}}

 \newcommand{\kim}[1]{\iffalse \textcolor{blue}{\textbf{[Kim: #1]}}\fi}
 \newcommand{\jesse}[1]{\iffalse \textcolor{red}{\textbf{[Jesse: #1]}}\fi}

\begin{document}

\maketitle 

\section{Abstract}
{
\bf
The effectiveness of Reinforcement Learning (RL) depends on an animal's ability to assign credit for rewards to the appropriate preceding stimuli. 
One aspect of understanding the neural underpinnings of this process involves understanding what sorts of stimulus representations support generalisation. 
The Successor Representation (SR), which enforces generalisation over states that predict similar outcomes, has become an increasingly popular model in this space of inquiries. 
Another dimension of credit assignment involves understanding how animals handle uncertainty about learned associations, using probabilistic methods such as Kalman Temporal Differences (KTD).
Combining these approaches, we propose using KTD to estimate a distribution over the SR.
KTD-SR captures uncertainty about the estimated SR as well as covariances between different long-term predictions.
We show that because of this, KTD-SR exhibits partial transition revaluation as humans do in this experiment without additional replay, unlike the standard TD-SR algorithm.
We conclude by discussing future applications of the KTD-SR as a model of the interaction between predictive and probabilistic animal reasoning.
}
\begin{quote}
\small
\textbf{Keywords:} 
Reinforcement Learning; Successor Representation; Kalman Filter; Transition Revaluation
\end{quote}

\section{Introduction}


An impressive signature of animal behavior is the capacity to flexibly learn relationships between the environment and reward.
One approach to understanding this behavior involves investigating how the brain represents different stimuli such that credit for reward is generalised appropriately.
Predictive representations, like the Successor Representation (SR) \cite{Dayan1993ImprovingRepresentation}, generalise over stimuli that predict similar futures and can provide a useful balance between efficiency and flexibility \cite{Gershman2018TheSubstrates, Russek2016a}. SR learning is faster to adapt to change than model-free (MF) learning, particularly changes in reward location, and supports more efficient state evaluation than model-based (MB) algorithms, which use time-consuming forward simulations to evaluate state. Since this efficiency depends on caching long-term expected state occupancies, however, the SR is worse than MB at handling changes in the environment's transition structure. In neuroscience and psychology, the SR offers a compelling explanation for a range of behavioural and neural findings \cite{Momennejad2016, Stachenfeld2017TheMap, Gardner2018RethinkingError, Garvert2017}.

While the SR offers a solution to some of the shortcomings of model-free learning, existing methods for estimating the SR, such as temporal difference (TD) learning, do not take into account uncertainty. Here, we attempt to rectify this by drawing on the Kalman TD (KTD) method for value learning \cite{Geist2010KalmanDifferences}, which explains a range of animal conditioning phenomena that standard TD cannot explain \cite{Gershman2015ALearning}. KTD-SR gives the agent an estimate of its uncertainty in the SR as well as the covariance between different entries of the SR.
We show how this augments the SRs capacity to support revaluation following changes in transition structure.

\section{Results}

\subsection{The successor representation}

We define an RL environment to be a Markov Decision Process consisting of \emph{states} $s$ the agent can occupy, \emph{transition probabilities} $T_\pi(s'|s)$ of moving from state $s$ to states $s'$ given the agent's policy $\pi(a|s)$ over actions $a$, and the reward available at each state, for which $R(s)$ denotes the expectation. An RL agent is tasked with finding a policy that maximises its expected discounted total future reward, or \emph{value}: 
\begin{equation}\label{eq:value_def}
    V(s)=\E_{\pi} \left[\sum_{t=0}^{\infty} \gamma^{t} R\left(s_{t}\right) | s_{0}=s\right]
\end{equation}
where $t$ indexes timestep and $\gamma$, where $0\leq\gamma<1$, is a discount factor that down-weights distal rewards.

The value function can be decomposed into a product of the reward function $R$ and the SR matrix $M$ \cite{Dayan1993ImprovingRepresentation}:
\begin{equation}\label{eq:factorised_value}
    V(s)=\sum_{s^{\prime}} M\left(s, s^{\prime}\right) R\left(s^{\prime}\right)
\end{equation}
$M$ is defined such that each entry $M(s,s')$ gives the expected discounted future number of times the agent will visit $s'$ from starting state $s$, under the current policy \cite{Dayan1993ImprovingRepresentation}:
\begin{equation}\label{eq:sr_def}
    M(s, s') = \E_{\pi} \left[ \sum_{t=0}^{\infty} \gamma^t \I(s_t = s') | s_0 = s \right]
\end{equation}
where $\I(s_t = s') =1$ if $s_t=s'$ and 0 otherwise. Each row $M(s, :)$ in this matrix constitutes the SR for some state $s$, thus representing each state as a vector over future ``successor states.''
Factorising value into an SR term and a reward term permits greater flexibility because if one term changes, it can be relearned while the other remains intact \cite{Dayan1993ImprovingRepresentation, Gershman2018TheSubstrates}.


We first consider the SR in a tabular setting with deterministic transitions and a fixed, deterministic policy.
This means that there is only one possible state $s_{t+1}$ following any predecessor state $s$. In this setting, the SR matrix rows of two temporally adjacent states $s_t, s_{t+1}$ can be recursively related as follows: 
\begin{equation}\label{eq:sr_deterministic}
    M(s_t,:) = \bm{\phi} (s)^T + \gamma M(s_{t+1},:),
\end{equation}
where $\bm{\phi} (s)$ is the feature vector (of length $n$, the number of features) observed by the agent in state $s$. 
In this article, we consider problems with discrete state spaces, for which the feature vector $\bm{\phi} (s)$ is a one-hot vector with an entry for every state and a 1 only in the \emph{s\textsuperscript{th}} position.
Equation \ref{eq:sr_deterministic} is analogous to the Bellman equation for value widely used in RL \cite{Sutton1998ReinforcementIntroduction}, with the vector-valued $M(s_t, :)$ in lieu of scalar $V(s_t)$.

We can express the estimated current one hot state vector (based on the SR) as the difference between two successive SRs: 
\begin{align}
    \hat{\bm{\phi}}(s_t) &= \hat{M}(s_t,:)^T - \gamma \hat{M}(s_{t+1},:)^T \nonumber \\
    &= \hat{M}^T \bm{\phi}(s_t) - \gamma  \hat{M}^T \bm{\phi}(s_{t+1}) \nonumber \\
    &= \hat{M}^T \bm{h}_t \label{eq:obs}
\end{align}
where we have defined $\mathbf{h}_t = \bm{\phi}(s_t) - \gamma \bm{\phi}(s_{t+1})$: the discounted temporal difference between state features. The (vector valued) successor prediction error, used to update the SR in TD methods, is then given by $\bm{\delta}_t = \bm{\phi}(s_t) - \hat{\bm{\phi}}(s_t)$.


\subsection{Learning a probabilistic SR using a Kalman Filter}

The algorithm described above produces a point estimate of the SR. While useful for approximating expected value, it is not capable of expressing certainty in these estimates. 
In order to derive a probabilistic interpretation of the SR, we assume that the agent has an internal generative model of how sensory data are generated from the SR parameters that can be learned with KTD \cite{Geist2010KalmanDifferences, Gershman2015ALearning}.
This model consists of a \emph{prior distribution} on the (hidden) parameters, $p(\bm{m}_0)$ -- where $\mathbf{m}_t = vec(M_t^T)$ is the SR reshaped into a vector -- an \emph{evolution process} on the parameters, $p(\bm{m}_t | \bm{m}_{t-1} )$, and a distribution of observed (one-hot) feature vectors given the current parameters and observations $p(\bm{\phi}_t | \bm{m}_{t}, \bm{h}_{t} )$. As with earlier work on KTD, we assume a Gaussian model: $\bm{m}_0 \sim \mathcal{N} \left( \bm{0}, C_{0|0} \right)$, $\bm{m}_t  \sim \mathcal{N} \left( \bm{m}_{t-1}, C_{v_t} \right)$ and $\bm{\phi}_t \sim \mathcal{N} \left( \hat{\bm{\phi}}_t,  C_{n_t} \right)$, 
where $C_{0|0}$ is the prior covariance between SR matrix entries, $C_{v_t}$ is the process covariance, describing how the evolution of different parameters covaries, and $C_{n_t}$ is the observation covariance, describing covariance in the observations. $C_{0|0}$, $C_{v_t}$ and $C_{n_t}$ are set by the practitioner (see Table \ref{tab:params}).  

The purpose of the Kalman Filter is to infer a posterior distribution over that hidden state $\bm{m}_t$ given the observations $\bm{\phi}$:
\begin{equation}\label{eq:posterior}
    p(\bm{m}_t | \bm{\phi}_{1:t} ) \propto p(\bm{\phi}_{1:t} |\bm{m}_t ) p(\bm{m}_t)
\end{equation}

Under the Gaussian model described above, this posterior distribution is Gaussian with mean $\bm{m}_t$ and covariance $C_t $ parameters which will be estimated by the Kalman Filter. To set up the filter, we specify an \emph{evolution equation} describing how the hidden parameters (the SR) evolve over time and an \emph{observation equation} describing how observation relates to our hidden parameters. These two equations comprise the \emph{state-space formulation} for KTD SR:
\begin{equation}
    \label{eq:linearstatespace}
    \begin{cases}
    \mathbf{m}_t = \mathbf{m}_{t-1} + \bm{v}_t & \text{(evolution equation)} \\
    \bm{\phi}(s_t) = ( \mathbf{h}_t \otimes I)^T \mathbf{m}_t + \bm{n}_t & \text{(observation equation)} 
    \end{cases}
\end{equation}
where $\bm{v}_t$ is the \emph{process noise} and $\bm{n}_t$ the \emph{observation noise}, $\otimes$ denotes the Kronecker product and $I$ the identity matrix. We will start from the assumption that the process noise is white, meaning that $\E[{\bm{m}_t}] = {\bm{m}_{t-1}}$, i.e. the expected mean SR on time $t$ equals the estimated SR on time $t-1$.

The Kalman Filter keeps track of the mean $\bm{m}_t$ and covariance $C_t$ of the posterior (\ref{eq:posterior}). At each timestep, the parameters of the posterior are updated using the Kalman Filter equations: 
\begin{align}
    \hat{\mathbf{m}}_{t|t} &= \hat{\mathbf{m}}_{t|t-1} + K_t (\bm{\phi}_t - \hat{\bm{\phi}}_t) \\
    C_{t|t} &= C_{t|t-1} - K_t C_{\bm{\phi}_t} K_t^T \\
    K_t &= C_{\mathbf{m}\phi_t} C_{\phi_t}^{-1}
\end{align}
where $C_{\mathbf{m}\phi_t}$ is the covariance between the parameters and the prediction error, and $C_{\phi_t}$ is the covariance of the prediction error. The notation $C_{t|t} = \E \left[C_t | \bm{\phi}_1 ... \bm{\phi}_t \right]$ means that the estimate of the parameter covariance is conditioned on all observations until time $t$ \cite<see>{Geist2010KalmanDifferences}.
Importantly, and in contrast to standard TD updates for the SR \cite{Dayan1993ImprovingRepresentation}, the Kalman gain $K_t$ is stimulus specific, as it is dependent on the ratio between the covariance in the parameters and the covariance in the observations. This allows for a principled weighting of prior knowledge and incoming data.
Note also that the Kalman gain is a matrix, meaning that each entry of the successor representation gets updated with a different gain.
See Algorithm \ref{alg:linearSRTD} for a full description of the method, including how these quantities are computed. 

\begin{algorithm}
\caption{Kalman TD Successor Representation}\label{alg:linearSRTD}
\SetAlgoLined
 Initialization: priors $\mathbf{m}_{0|0}$ and $C_{0|0} $ \;
 \For{$t \leftarrow 1,2,...$}{
  Observe transition $(s_t, s_{t+1})$ \;
  \emph{Prediction step}\;
  $\mathbf{\hat{m}}_{t|t-1} = \mathbf{\hat{m}}_{t-1|t-1}$ \;
  $C_{t|t-1} = C_{t-1|t-1} + C_{v_{t}} $ \;
  \emph{Compute statistics of interest} \; 
  $ \hat{\bm{\phi}}(s_t)  = (\mathbf{h}_t \otimes I)^T \hat{\mathbf{m}}_t $ \;
  $C_{\mathbf{m}\bm{\phi}_t} = C_{t|t-1} (\mathbf{h}_t \otimes I)$ \;
  $C_{\phi_t} = (\mathbf{h}_t \otimes I)^T C_{t|t-1} (\mathbf{h}_t \otimes I) + C_{n_t}$ \;
  \emph{Correction step} \; 
  $K_t = C_{\mathbf{m}\phi_t} C_{\phi_t}^{-1}$ \;
  $\hat{\mathbf{m}}_{t|t} = \hat{\mathbf{m}}_{t|t-1} + K_t (\bm{\phi}_t - \hat{\bm{\phi}}_t)$ \;
  $C_{t|t} = C_{t|t-1} - K_t C_{\bm{\phi}_t} K_t^T$
  }
\end{algorithm}

In summary, we have introduced a method of handling uncertainty over SR estimates. This allows for an efficient combination of prior knowledge and incoming information when updating the SR estimates. Furthermore, it allows us to estimate dependencies between different entries in the SR that inform SR updates. This permits non-local updates which, in the case of KTD for value estimation, have proven to better explain animal behaviour than the strictly local updates of vanilla TD \cite{Gershman2015ALearning}. We explore a possible role for non-local updates in the following section.

\subsection{Partial Transition Revaluation Simulations}

\begin{figure}[ht]
\begin{center}
    \includegraphics[width=.48\textwidth]{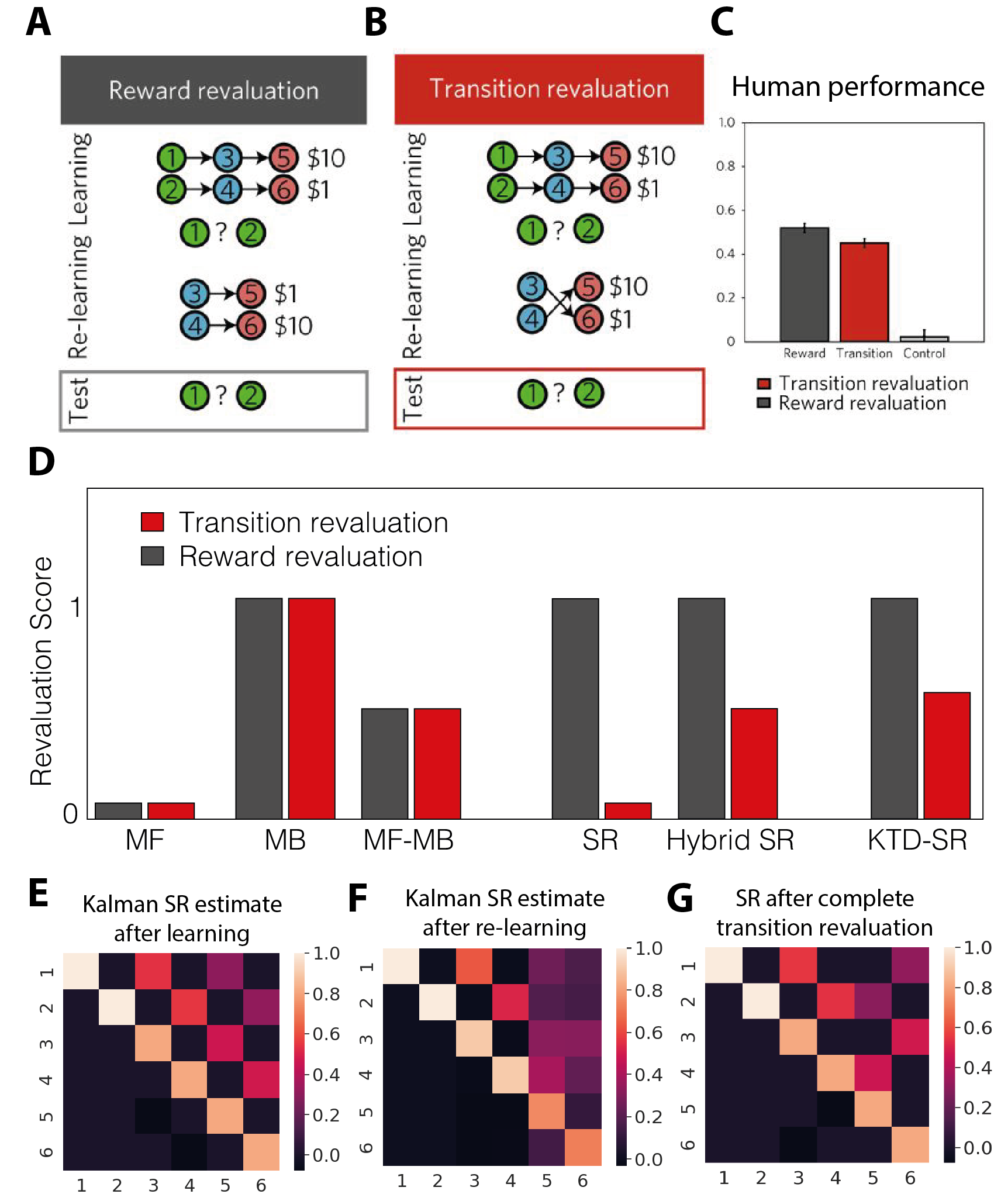}
\end{center}
\caption{KTD SR performance on a transition and reward revaluation experiment. 
(A-B) Task structure for (A) reward revaluation and (B) transition revaluation experiments. (C) Human performance on transition and reward revaluation tasks. (D) Model predictions for classic model-free, model-based or a hybrid of model free and model-based algorithms, TD-SR, hybrid SR and KTD-SR. (E-G) The SR matrix estimated by KTD after (E) learning (phase 1), (F) re-learning (phase 2) and (G) after a hypothetical complete transition revaluation. \emph{Panels A--C reprinted with permission from \shortciteA{Momennejad2016}.}}
\label{fig:transition}
\end{figure}

A key prediction of standard TD-SR learning is that ``reward revaluation'' should be supported while ``transition revaluation'' should not.
\shortciteA{Momennejad2016} tested this in humans.
In the first phase of their experiment, participants learned two different sequences of states terminating in different reward amounts: 2$\rightarrow$4$\rightarrow$6$\rightarrow$\$1 and 1$\rightarrow$3$\rightarrow$5$\rightarrow$\$10 (see Figure \ref{fig:transition}B).
In the next stage, half of the participants were exposed to the transition revaluation condition, observing novel 4$\rightarrow$5$\rightarrow$\$10 and 3$\rightarrow$6$\rightarrow$\$1 transitions. 
The other half experienced ``reward revaluation'' in the form of novel reward amounts 6$\rightarrow$\$10 and 5$\rightarrow$\$1 (Figure \ref{fig:transition}A). 
Importantly, the novel experiences start from intermediate states 
such that transitions from 1 or 2 are not seen following phase 1. While participants were significantly better at reward revaluation than transition revaluation, they were capable of some transition revaluation as well (Figure \ref{fig:transition}C). Accordingly, the authors proposed a hybrid SR model: an SR-TD agent that is also endowed with capacity for replaying experienced transitions (Figure \ref{fig:transition}F). This permits updating of the SR vectors of states 1 and 2 through simulated experience.
\begin{table}[]
\caption{Parameter values}
\begin{tabular}{@{}lll@{}}
\toprule
Name        & Symbol            & Value             \\ \midrule
Discount factor    & $\gamma$          & 0.9               \\
Process covariance & $C_{v_{t}}$     & $(1 \times 10^{-3}) I$          \\
Observation covariance & $C_{n_t}$       & $I$ \\
Prior covariance & $C_{0|0}$       & $0.1 I$           \\
Prior SR & $\bm{m}_{0|0}$  & $vec(I)$          \\
Rescorla Wagner learning rate & $\alpha_r$        & 0.1               \\ 
Number of trials per phase & $N$        & 50               \\ \bottomrule
\end{tabular}
\label{tab:params}
\end{table}

Here, we simulate this experiment and find that the probabilistic KTD-SR accounts for partial transition revaluation even without replay (Figure \ref{fig:transition}D). 
KTD-SR correctly learns the SR matrix after phase 1 (Figure \ref{fig:transition}E) as well as an estimate of the covariance between all entries in the SR matrix, $C_{t|t}$. 
Unlike TD-SR, KTD-SR uses the covariance matrix to estimate the Kalman gain and uses that to update the whole matrix. This means that after seeing $3\rightarrow6$, it updates not just $M(3, :)$ but also $M(1, 6)$ because these entries have historically covaried (same for $M(4, :)$ and $M(2,5)$) (Figure \ref{fig:transition}F). 
To estimate direct reward $\hat{r}$, the agent uses a Rescorla-Wagner rule \cite{Rescorla1972ANonreinforcement}.
Model parameters are listed in Table \ref{tab:params} and experimental parameters are kept the same as in \cite{Momennejad2016}.


\section{Discussion}

The SR constitutes a middle ground between model-based and model-free RL algorithms by separating reward representations from cached long-run state predictions. Here we learn a probabilistic SR model using KTD that supports principled handling of uncertainty about state predictions and inter-dependencies between these predictions. We exploit this feature to show that, unlike standard TD-SR, KTD-SR can perform partial transition revaluation. 
In later work, we plan to test our model on other tasks that could benefit from KTD-SR in a similar way, such as policy revaluation \cite<a well-known weak spot of TD-SR; >{Barreto2016}. 

We note the relative strengths and weaknesses of KTD-SR when compared to a hybrid-MB-SR approach.
Replay requires a buffer to store experienced episodes and a sufficient number of replays that information is propagated throughout the SR model. 
While KTD-SR can incorporate information about long-range in a single update, it must learn and store a large $n^2\times n^2$ matrix  \cite<although dimensionality reduction can reduce this burden;>{Fisher1998DevelopmentFilter}. 
There is compelling evidence in favor of both replay \cite{Carr2011HippocampalRetrieval, Olafsdottir2018ThePlanning} and probabilistic representations \cite{Ma2006BayesianCodes} driving behavior.
Future work will consider how the relative tradeoffs of these approaches constrain hypotheses.

Probabilistic models provide a number of advantages for RL in terms of optimal credit assignment \cite{Kruschke2008BayesianLearning}, uncertainty-minimising exploration \cite{Dearden1998BayesianQ-learning}, arbitration between competing models \cite{Daw2005Uncertainty-basedControl}. Distributional RL-trained neural network agents achieve state of the art performance \cite{Bellemare2017ALearning}.
Furthermore, a range of animal learning findings suggest that animals are capable of probabilistic reasoning \cite{Gershman2015ALearning, Kruschke2008BayesianLearning,Courville2006BayesianWorld}.
Future work will involve exploring these advantages in the context of SR learning \cite{Gardner2018RethinkingError}.

We make several assumptions in order to make this model tractable.
The Gaussian assumption is clearly violated in the case of one-hot state vectors (i.e. neither $\phi$ nor $M$ should have negative entries). However, the model is sufficiently expressive that a good approximation can still be found, and a ``successor feature'' model could be applied over arbitrary features for which the Gaussian assumption might hold.
The random walk process noise is useful for capturing slow changes in the environment, but might be ill-suited for step changes or sub-optimal when the dynamics are predictable. 
While we assume deterministic transitions and linear function approximation in this work, it is straightforward to extent the model to stochastic transitions and nonlinear function approximation with a ``coloured noise'' approach \cite{Geist2010KalmanDifferences}.

\section{Acknowledgments}
This work is funded by the Gatsby Foundation and the Wellcome Trust. We thank Samuel Gershman, Talfan Evans, Eszter V\'{e}rtes, Steven Hansen and Matthew Botvinick for helpful comments and suggestions.

\bibliographystyle{apacite}

\setlength{\bibleftmargin}{.125in}
\setlength{\bibindent}{-\bibleftmargin}

\bibliography{references.bib}

\end{document}